\documentclass[aip,reprint]{revtex4-1}

\usepackage{color}
\usepackage{graphicx}
\usepackage{amsmath}
\usepackage{subcaption}

\usepackage{natbib}

\begin{document}
\title{Non-Linear Ablative Rayleigh-Taylor Instability: Increased Growth due to Self-Generated Magnetic Fields}

\author{C. A. Walsh}
\email{walsh34@llnl.gov}
\affiliation{Lawrence Livermore National Laboratory}
\author{D. S. Clark}
\affiliation{Lawrence Livermore National Laboratory}

\date{\today}

	\begin{abstract}
	The growth rate of the non-linear ablative Rayleigh-Taylor (RT) instability is enhanced by magnetic fields self-generated by the Biermann battery mechanism; a scaling for this effect with perturbation height and wavelength is proposed and validated with extended-magnetohydrodynamic simulations. The magnetic flux generation rate around a single RT spike is found to scale with the spike height. The Hall Parameter, which quantifies electron magnetization, is found to be strongly enhanced for short wavelength spikes due to Nernst compression of the magnetic field at the spike tip. The impact of the magnetic field on spike growth is through both the suppressed thermal conduction into the unstable spike and the Righi-Leduc heat-flow deflecting heat from the spike tip to the base. Righi-Leduc is found to be the dominant effect for small Hall Parameters, while suppressed thermal conduction dominates for large Hall Parameters. These results demonstrate the importance of considering magnetic fields in all perturbed inertial confinement fusion hot-spots.
	\end{abstract}
	\maketitle
	
	\begin{figure*}
		\centering
		\centering
		\includegraphics[scale=0.7]{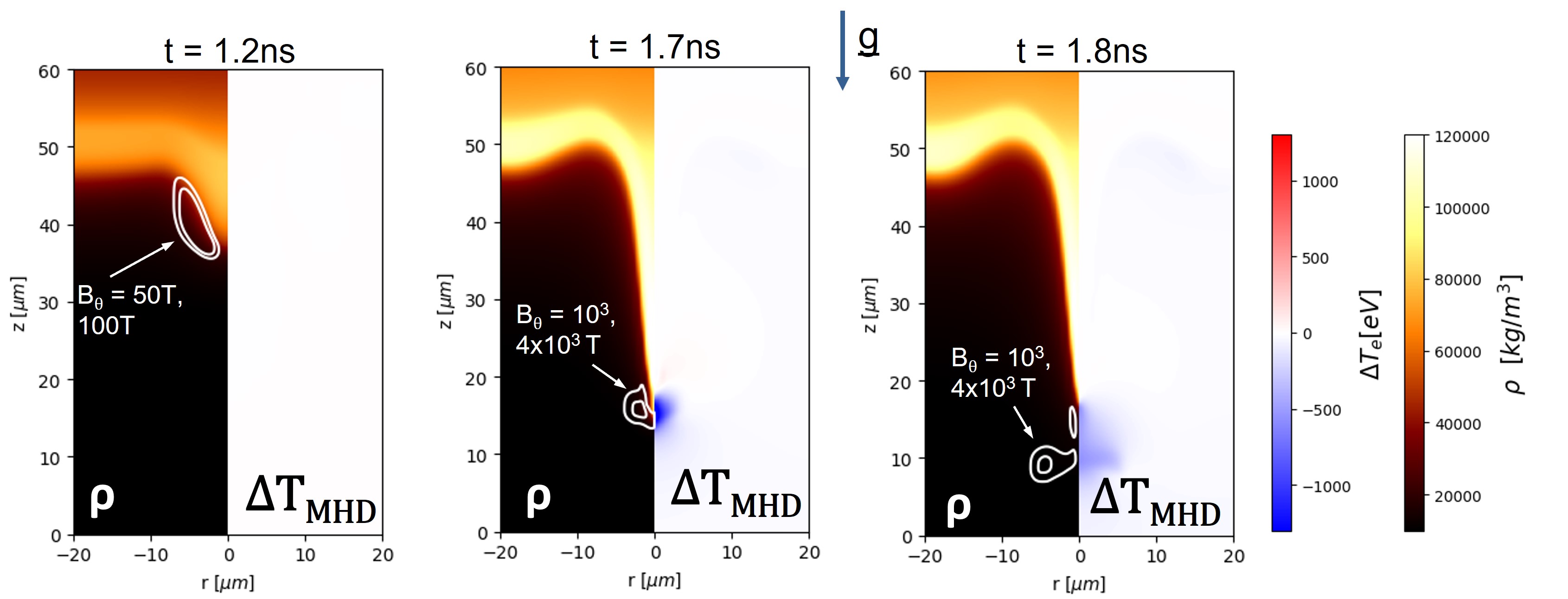}\caption{ \label{fig:Density_delTe} Density and change in temperature due to MHD effects at 3 different times during the ARTI evolution for a perturbation of $20\mu$m wavelength. Select magnetic field strength contours have been indicated.}	
	\end{figure*}
	
	The Rayleigh Taylor instability\cite{KULL1991197,zhou2017,zhou2019} (RTI) results in spike-bubble growth when a light fluid is accelerated into a dense fluid. The instability is prevalent in astrophysics\cite{1996ApJ...456..225H,cabot2006,Calder_2002,Arnett_2000} as well as laboratory plasma experiments, such as the acceleration\cite{PhysRevLett.33.761,hsing1997,raman2014,Smalyuk_2012,Smalyuk_2019} and deceleration \cite{clark2016,clark2019,betti2001,weber2015a} phases of inertial confinement fusion (ICF) implosions. In ICF experiments, the RTI is seeded by target imperfections \cite{casey2021,gatujohnson2020,smalyuk2015,mcglinchey2018,hammel2015} or asymmetries in the drive \cite{li2004,gatujohnson2019,spears2014}. RTI growth is lowered when a transfer of energy (for example electron heat-flow or radiation) results in ablation of the dense spikes, which is called the ablative Rayleigh-Taylor instability (ARTI) \cite{PhysRevLett.89.195002,PhysRevLett.76.4536,betti2001,betti1995,betti1998,kilkenny1994}.
	
	Magnetic fields are anticipated to be self-generated at perturbations by the Biermann battery mechanism, both during acceleration of the ablator \cite{manuel2012,PhysRevLett.110.185003,manuel2015,hill2017,sadler2021} and during the deceleration of the fusion fuel \cite{walsh2017,srinivasan2012a,walsh2021a}. Magnetic fields up to 10kT are expected in the hot-spot, which is large enough to magnetize the electron population \cite{walsh2017}. Previous ICF capsule simulations demonstrated that electron magnetization could enhance the growth of a perturbation by $7\mu m$ within a hot-spot of $30\mu m$ radius \cite{2021}. While a theory for self-generated magnetic fields has been developed for the linear stage relevant to the acceleration phase of ICF implosions \cite{garcia-rubio2021}, the understanding of how magnetic fields affect different wavelengths or amplitudes of ARTI in the highly non-linear deceleration phase remains lacking.
	
	This letter proposes an improved theory for the ARTI that includes the effect of self-generated magnetic fields. The theory is derived in 3 stages: magnetic flux quantification; magnetic flux concentration at spike tips; thermal conduction magnetization. The change to the ARTI spike velocity due to xMHD will be demonstrated to be:
	
	\begin{equation}
		\Delta V_{MHD} \sim V_{abl} \frac{\kappa_{\parallel}^c + \kappa_{\wedge}^c - \kappa_{\bot}^c}{\kappa_{\parallel}^c} \label{eq:delta_V}
	\end{equation}
	
	Where $V_{abl}$ is the ablation velocity of cold fuel due to electron thermal conduction. The $\kappa^c$ coefficients are thermal conductivity components \cite{epperlein1986}, depending on the plasma ionization ($Z$) and the electron Hall Parameter ($\omega_e \tau_e$). It will be shown that the scaling for $\omega_e \tau_e$ with spike height ($h$) and wavelength ($\lambda$) is:
	
	\begin{equation}
		\omega_e \tau_e \sim \frac{h \int h \delta t}{\lambda^2} \label{eq:wt}
	\end{equation}
	
	Simulations of a simplified ARTI test problem demonstrate the validity of the proposed theory. For this, the xMHD Gorgon code \cite{ciardi2007,chittenden2009,walsh2017} is used. Gorgon includes magnetic fields self-generated through the Biermann battery process \cite{walsh2017,PhysRevLett.125.145001} as well as through composition gradients \cite{sadler2020}. Kinetic suppression of the Biermann battery generation rate is included \cite{sherlock2020,campbell2021measuring}, although the results of this paper are too collisional to be affected by this term. The transport of magnetic fields includes bulk plasma advection, Nernst, cross-gradient-Nernst and resistive effects \cite{walsh2020}. In this letter the magnetic field only feeds back on the plasma through magnetization of the electrons, as the magnetic pressure is insignificant. Magnetization of the electrons uses an anisotropic thermal conduction algorithm \cite{sharma2007}, including Righi-Leduc heat-flow \cite{walsh2018a}. Updated magnetized transport coefficients are used \cite{sadler2021,davies2021}, which have been shown to affect spike propagation \cite{2021}.
	
	The ARTI test problem is the same as has been used previously to investigate the impact of an applied magnetic field on instability growth \cite{walsh2021magnetized}. A light ($\rho_{L0}=10^3$kg/m$^3$) deuterium-tritium (DT) plasma is accelerated ($g=10^{15}$m/s$^2$) into a dense DT plasma ($\rho_{H0}=10^4$kg/m$^3$). The system is initialized with a transition layer ($d=200\mu$m) such that the density decays exponentially between $\rho_{H0}$ and $\rho_{L0}$. The low density DT plasma is initialized at $T_0 = 200$eV, with the temperature elsewhere set such that the system is isobaric. 2-D cylindrical ($r-z$) simulations are utilized, with the spike propagating down $z$. The boundary conditions chosen here are out-flowing. The grid resolution is kept at $0.5\mu$m for all cases. The choice of thermal and Nernst flux-limiters are not found to modify the spike propagation and are kept at a value of $0.1$.

	The system inputs ($g$, $\rho_{L0}$, $\rho_{H0}$, $T_0$, $d$) have been chosen to mimic hot-spot conditions, with the hot plasma reaching 2.4keV $40\mu$m from the peak density location ($\rho \approx 1.2\times 10^5$kg/m$^3$) at t=1.5ns. This temperature increases to 3keV by t=1.8ns, with the energy being sourced by external gravitation.
	
	A perturbation is applied at t=0 by offsetting the light-dense transition layer by $h_0$ along the acceleration axis. Throughout this letter, the 'small' perturbation case uses $h_0=0.2\mu m$ and the 'large' perturbation case uses $h_0=0.5\mu m$. The spike height $h$ is defined by the position of the $1$keV electron temperature contour.

	Figure \ref{fig:Density_delTe} shows the evolution for perturbation wavelength $\lambda = 20\mu m$ and large initial height. Contours of magnetic field strength are indicated over the plasma density, with the change in electron temperature due to the self-generated magnetic fields shown on the right hand side. At t=1.2ns the magnetic field has grown to over 100T, but there is no significant change to the case where MHD is ignored. At 1.7ns the magnetic field, which is primarily bunched around the spike tip, has grown to over 4kT; at this level the MHD has reduced the temperature near the spike tip by over 1keV. At 1.8ns the magnetic field has detached from the spike tip and is propagating into the hot plasma, a process called magnetic flux injection. Up until this time the impact of MHD continually increases. Once injection takes place the impact of MHD on spike growth decreases. The theory in this paper is only applicable while the magnetic field remains attached to the spike. 
	
	The first stage in the theory is quantifying the amount of magnetic flux around a perturbation. For a system with closed boundaries, the change in magnetic flux with time is \cite{walsh2021a}: 

\begin{equation}
	\frac{\partial \Phi_B}{\partial t} = \oint \frac{\nabla P_e}{e n_e} \cdot \delta \underline{l}
 - \oint \frac{\beta_{\parallel} \nabla T_e}{e} \cdot \delta \underline{l} \label{eq:flux}
 \end{equation}
 Where $\underline{l}$ is the path along the boundary. 
 
 The two terms in equation \ref{eq:flux} are sources of magnetic field. $\nabla P_e/en_e$ is the Biermann battery effect and $- \beta_{\parallel}\nabla T_e/e$ is a corrective term for gradients in $Z$ \cite{sadler2020}, although this is insignificant for the system of interest in this letter \cite{walsh2021a}.
 
 Previous work studied the growth of magnetic flux in capsule simulations due to the Biermann battery effect, finding the following relation for a single-mode case\cite{walsh2021a}:
 
 \begin{equation}
 	\frac{\partial \Phi_B}{\partial t} = \frac{T_h - T_c}{e} \ln \frac{\rho_c}{\rho_h} 
 	\Big( \frac{\Delta \rho R}{\rho R}  - \frac{\Delta T R}{T R}\Big) \label{eq:flux2}
 \end{equation}
 Where $\rho R$ is the line-integrated density, $TR$ is the line-integrated temperature and $\Delta$ is the single-mode variation due to the perturbation. $T_h$, $T_c$, $\rho_h$ and $\rho_c$ are bulk temperatures and densities of the hot and cold regions. This equation was found to compare favorably with full extended-MHD simulations of ICF hot-spots \cite{walsh2021a}.

Equation \ref{eq:flux2} can be simplified further by noting that $\Delta\rho R \approx h(\rho_c - \rho_h)$ and $\Delta T R \approx h(T_c - T_h)$, which gives a scaling for magnetic flux generation:

\begin{equation}
	\frac{\partial \Phi_B}{\partial t} \sim h \label{eq:scaling_a}
\end{equation}
i.e. the flux generation rate is independent of perturbation wavelength. Note that the domain size is changing when considering different perturbation wavelengths; if instead a whole capsule surface is considered, then more magnetic flux would result from the surface being filled with short wavelength features rather than long wavelength of the same amplitude. However, the flux per feature would be the same. 

\begin{figure}
	\centering
	\begin{subfigure}[b]{0.5\textwidth}
		\centering
		\includegraphics[width=1.\textwidth]{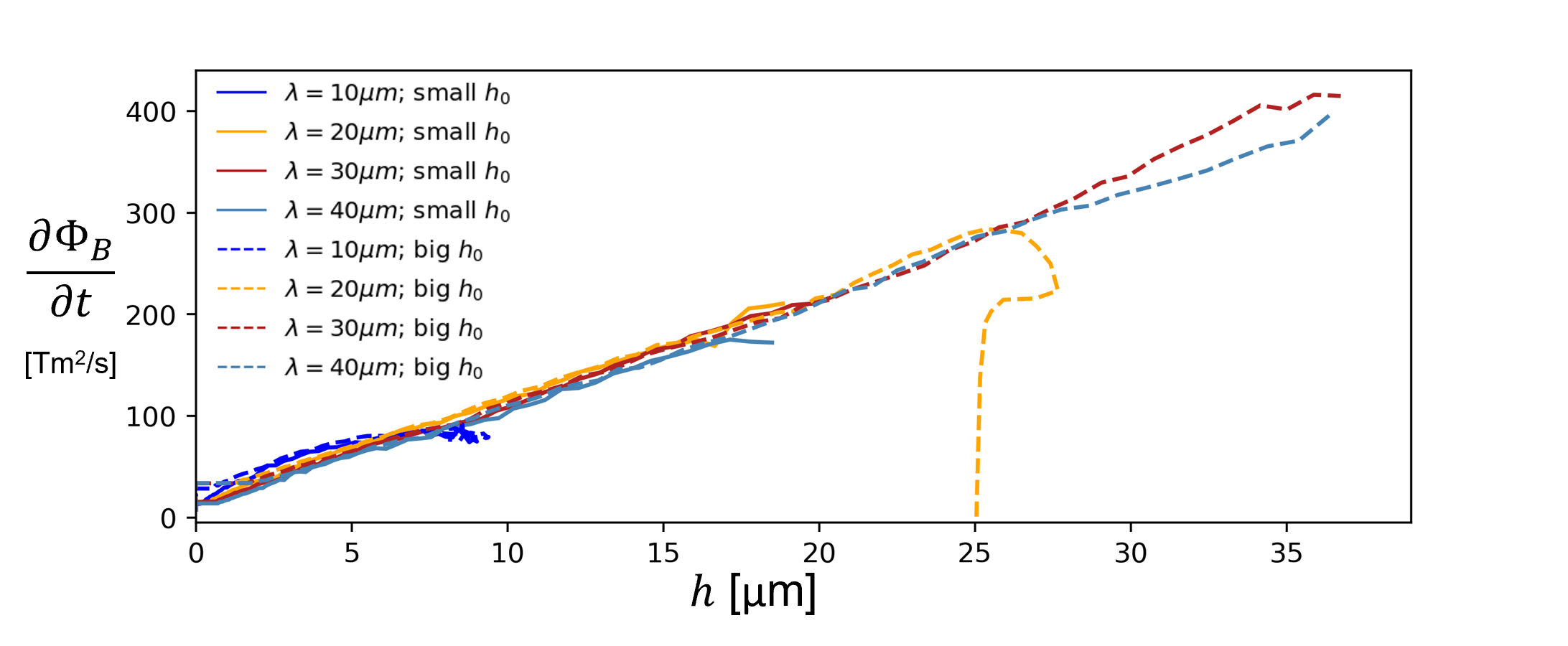}
		\caption{}
		\label{fig:scaling_a}
	\end{subfigure}
	\begin{subfigure}[b]{0.5\textwidth}
		\centering
		\includegraphics[width=1.\textwidth]{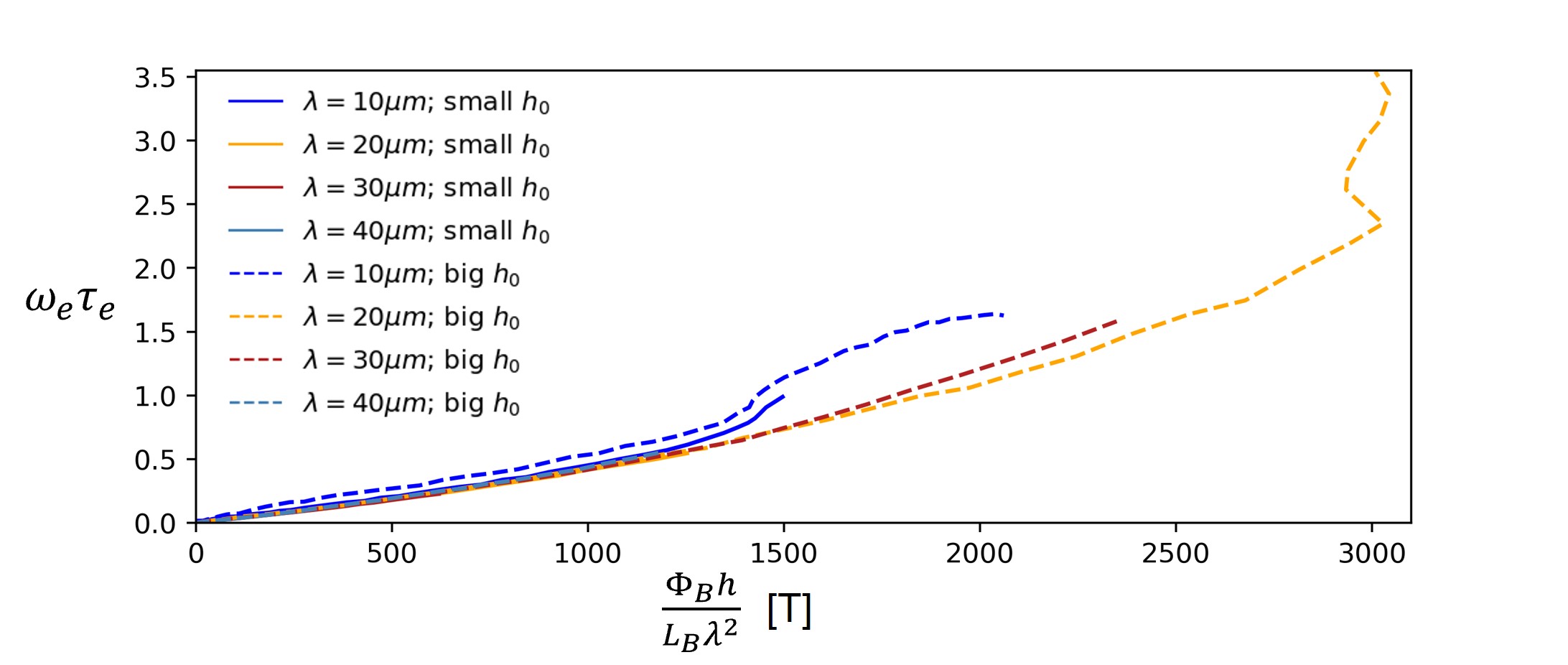}
		\caption{}
		\label{fig:scaling_b}
	\end{subfigure}
	\begin{subfigure}[b]{0.5\textwidth}
		\centering
		\includegraphics[width=1.\textwidth]{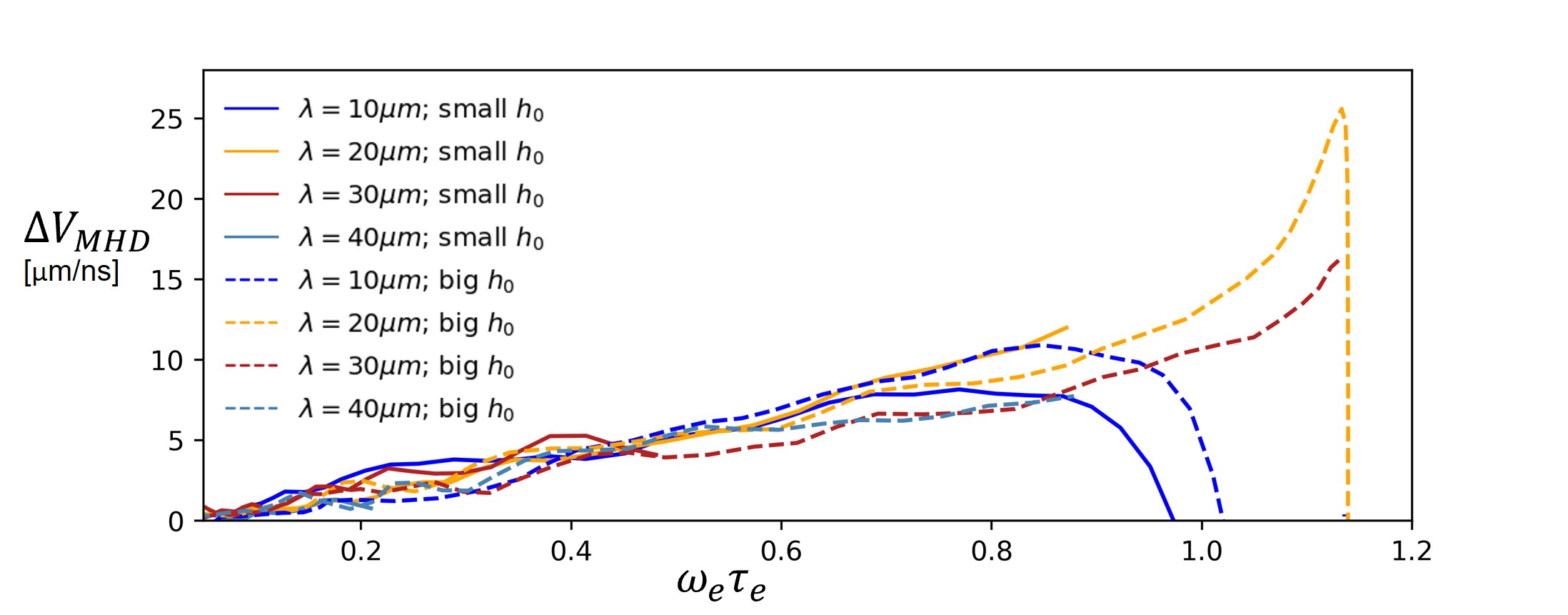}
		\caption{}
		\label{fig:scaling_c}
	\end{subfigure}
	
	\caption{These 3 figures demonstrate the effectiveness of the scalings proposed in this letter by comparison with xMHD simulations with different perturbation wavelengths and initial heights. (a) is for the magnetic flux generation (equation \ref{eq:scaling_a}); (b) is for the compression of magnetic flux at the spike tip (equation \ref{eq:scaling_b}); (c) is for the impact of magnetization on the spike velocity (equation \ref{eq:delta_V}).}
	\label{fig:scaling}
\end{figure}

Figure \ref{fig:scaling_a} plots the magnetic flux generation rate against spike height for simulations with different wavelengths and initial perturbation sizes, showing a good correlation until the magnetic flux injection regime is reached. Once the magnetic flux detaches from the spike, there can be significant resistive diffusion across the axis, resulting in the annihilation of flux. 

Next, theory is developed to show how the magnetic field is concentrated at the spike tip. The transport of magnetic fields in an extended-magnetohydrodynamic (xMHD) plasma is governed by the following equation \cite{walsh2020}: 

\begin{align}
\begin{split}
\frac{\partial \underline{B}}{\partial t} = & - \nabla \times \frac{\eta}{\mu_0 } \nabla \times \underline{B} + \nabla \times (\underline{v}_B \times \underline{B} ) \\
&+ \nabla \times \Bigg( \frac{\nabla P_e}{e n_e} - \frac{\beta_{\parallel} \nabla T_e}{e}\Bigg) \label{eq:mag_trans_new}
\end{split}
\end{align}

Where the first term on the right is resistive diffusion with diffusivity $\eta$ and the second term is advection of the magnetic field at velocity $\underline{v}_B$:

\begin{equation}
\label{eq:mag_trans_new_velocity}	\underline{v}_B = \underline{v} - \gamma_{\bot} \nabla T_e - \gamma_{\wedge}(\underline{\hat{b}} \times \nabla T_e) 
\end{equation}

Where the current-driven terms \cite{10.1088/1361-6587/ac3f25} have been neglected due to their insignificance in this regime \cite{walsh2020}. $\underline{v}$ is the bulk plasma velocity, while the $\gamma_{\bot}$ and $\gamma_{\wedge}$ terms are the Nernst and cross-gradient-Nernst terms respectively. $\underline{\hat{b}}$ is the magnetic field unit vector.

\begin{figure}
	\centering
	\centering
	\includegraphics[scale=0.6]{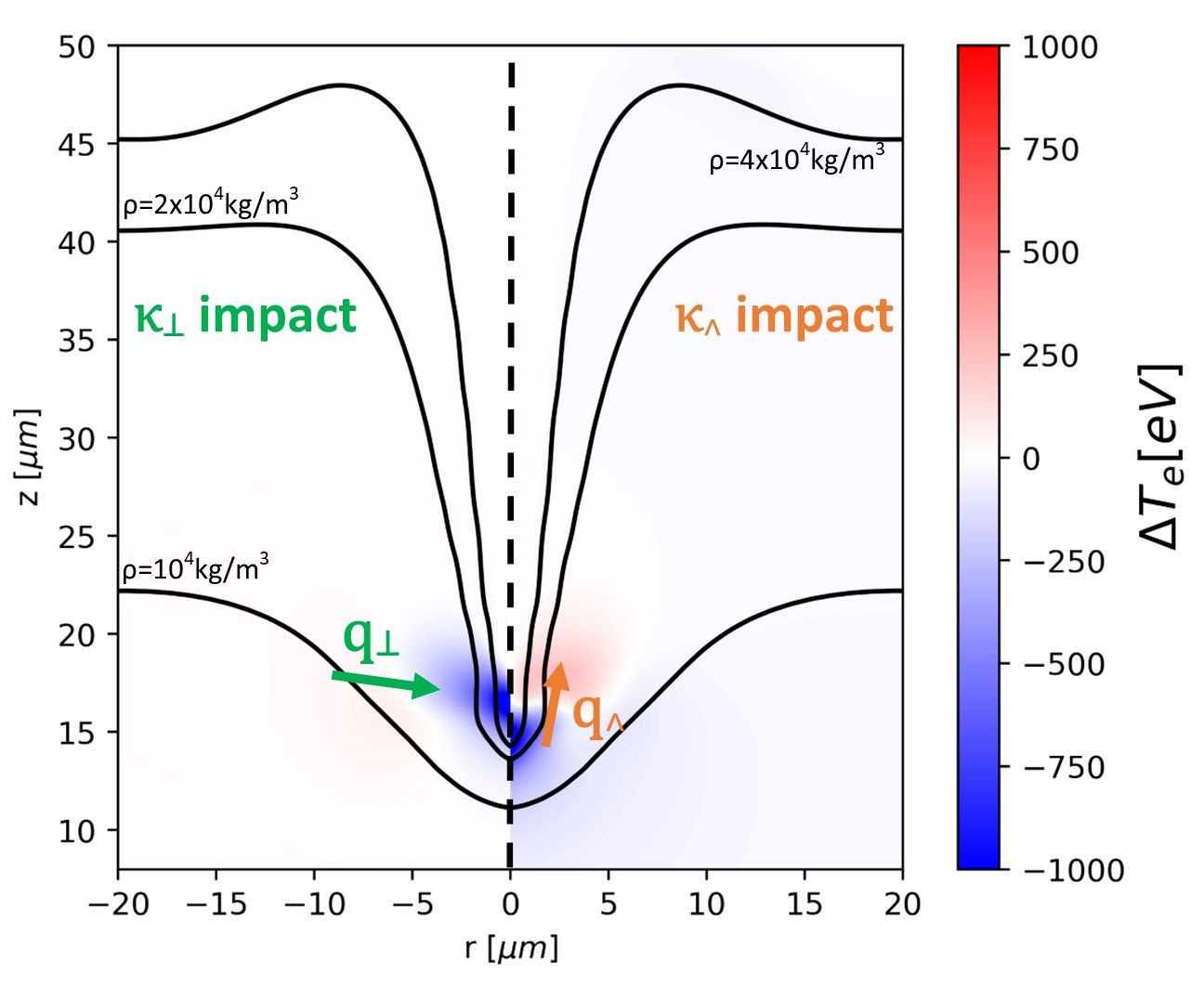}\caption{ Change in electron temperature of a $\lambda=20\mu$m spike due to magnetized thermal conduction (left) and Righi-Leduc heat-flow (right). The left-hand plot is the difference in temperature when just magnetized heat-flow ($\kappa_{\bot}$) is used. The plot on the right is the difference in temperature when Righi-Leduc heat-flow is also turned on. Density contours for the case including full MHD have been overlaid. \label{fig:diffTe_perp_wedge}}	
\end{figure}

If the magnetic flux was evenly distributed along the perturbation surface then the magnetic field strength would be approximately $\Phi_B/\lambda L_B$, where $L_B$ is the magnetic field length scale into the spike. $L_B$ is set primarily by Nernst advection of magnetic field along the acceleration axis from hot to cold plasma, which is independent of perturbation wavelength or amplitude; indeed, $L_B \approx 4\mu$m for all cases simulated. However, the peak field strength is found to be much larger than $\Phi_B/\lambda L_B$ for all cases and is dependent on the radial advection of magnetic field by Nernst into the spike tip. The ratio of radial Nernst advection (which compresses the flux) to the axial Nernst advection scales with $h/\lambda$. Assuming that this is the factor by which the magnetic flux compresses gives the electron magnetization scaling as:

\begin{equation}
	\omega_e \tau_e \sim \frac{\Phi_B h}{L_B \lambda^2} \label{eq:scaling_b}
\end{equation}

Where $\omega_e \tau_e \sim |\underline{B}| T_e^{3/2}/n_e$. Equation \ref{eq:scaling_b} is compared with simulations in figure \ref{fig:scaling_b}. While figure \ref{fig:scaling_a} shows that the magnetic flux generation rate decreases when the RT spikes reach the flux injection regime, the electron magnetization actually increases, as the magnetic fields are transported to higher temperature, lower density plasma that is easier to magnetize. The overall equation for thermal conduction magnetization (equation \ref{eq:wt}) is reached by combining equations \ref{eq:scaling_a} and \ref{eq:scaling_b}). Nernst has been confirmed as the cause of the additional $h/\lambda$ dependence in equation \ref{eq:scaling_b} by artificially reducing only the radial component, which results in the expected reduction to electron magnetization.

Finally, with an understanding developed for how the magnetic field bunches at the spike tip, it is possible to turn attention to how the magnetic field feeds back on the plasma hydrodynamics; this is not done directly through the magnetic pressure (the plasma $\beta$ is too large in all cases), but instead through magnetization of the electron population. Magnetized electron heat-flow follows:

\begin{figure}
	\centering
	\centering
	\includegraphics[scale=0.5]{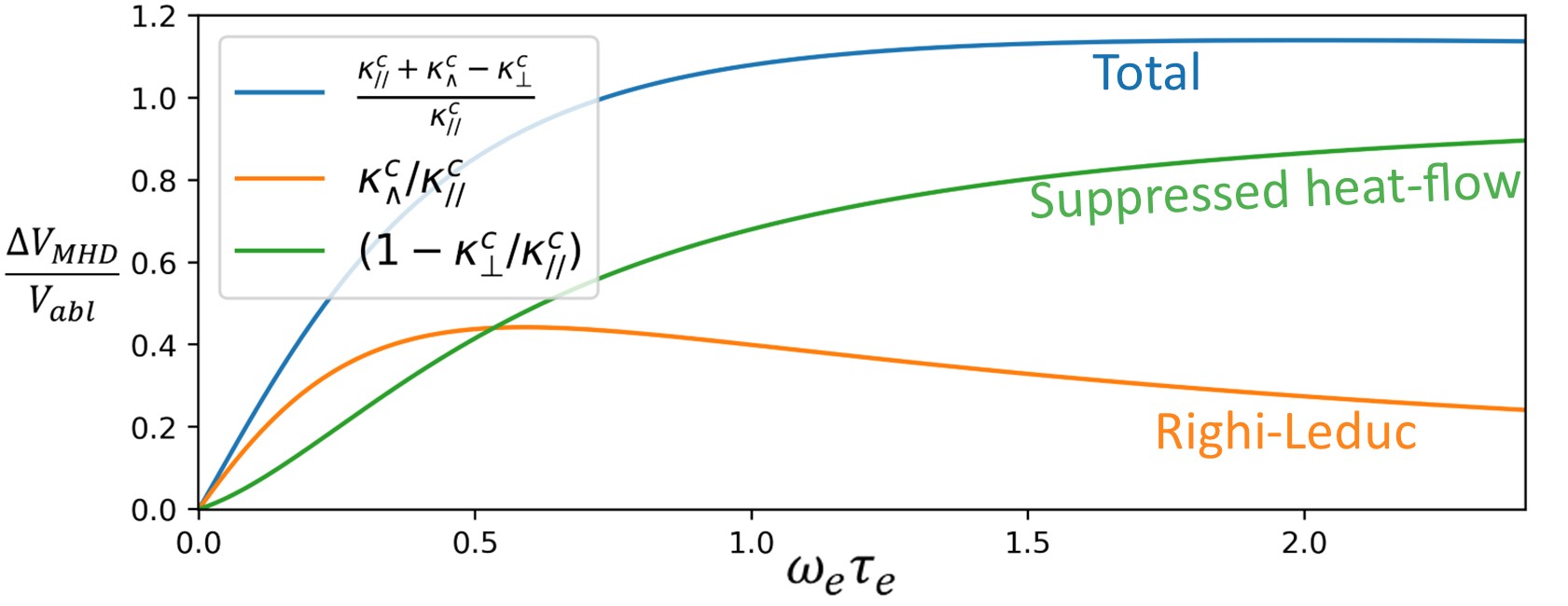}\caption{ \label{fig:kappa} Increase in RT spike velocity with electron magnetization, including separate curves showing the impact of suppressed perpendicular conduction and the Righi-Leduc component. }	
\end{figure}

\begin{figure}
	\centering
	\centering
	\includegraphics[scale=0.45]{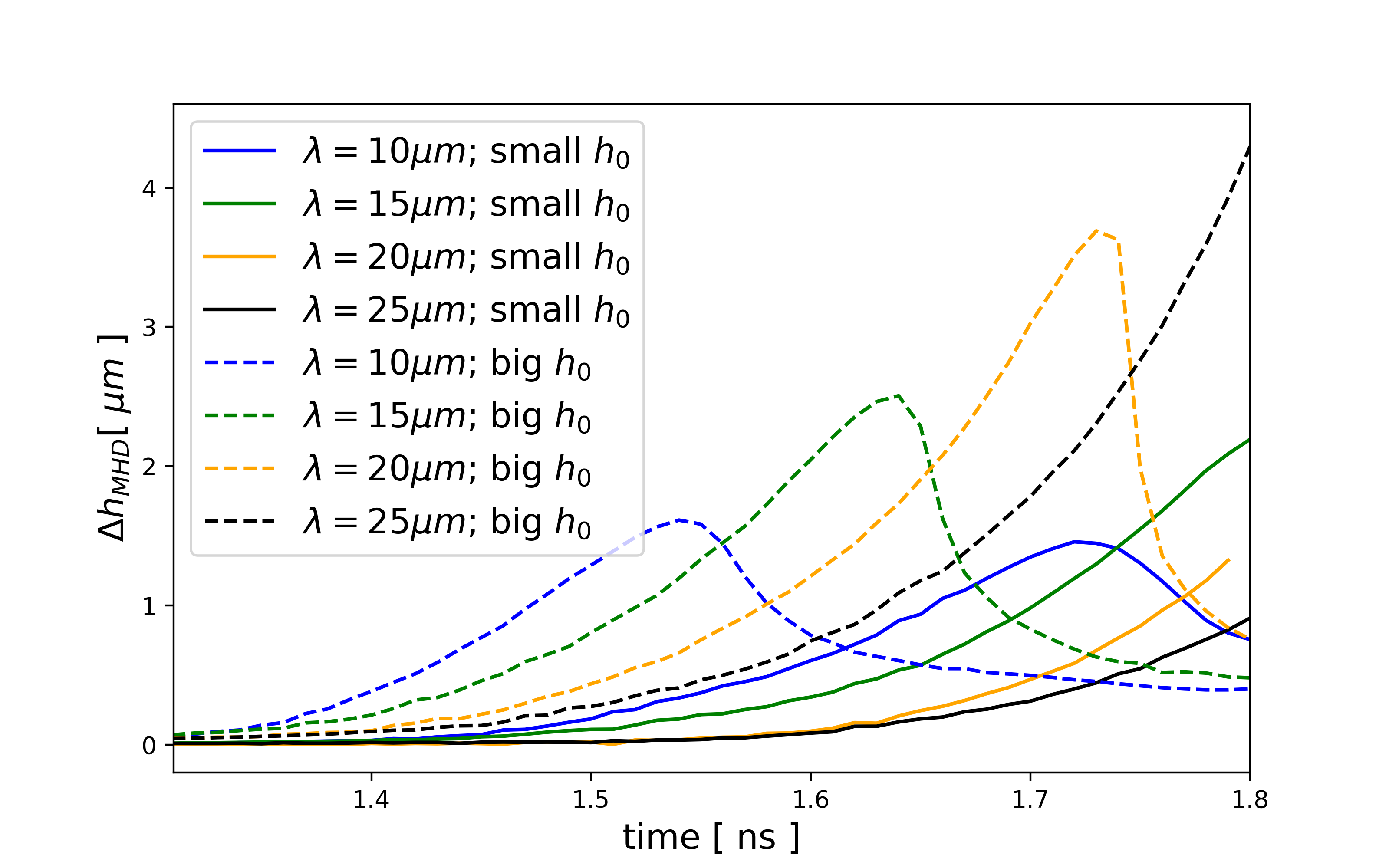}\caption{Increase in spike height versus time due to self-generated magnetic fields for a number of perturbation wavelengths and initialized heights. \label{fig:del_h}}	
\end{figure}

\begin{equation}
	\underline{q}_e = -\kappa_{\parallel} \nabla_{\parallel}T_e - -\kappa_{\bot} \nabla_{\bot}T_e - \kappa_{\wedge} \underline{\hat{b}} \times \nabla T_e 
\end{equation}
Where $\kappa_{\parallel}$ is the unmagnetized thermal conductivity along field lines; $\kappa_{\bot}$ is the thermal conductivity perpendicular to field lines that decreases with $\omega_e \tau_e$; $\kappa_{\wedge}$ is the Righi-Leduc coefficient. Righi-Leduc represents the electron heat-flow being re-directed 90 degrees due to the magnetic field and peaks for $\omega_e \tau_e \approx 0.5$ in a DT plasma \cite{epperlein1986}. Note the similarities between this equation and the magnetic field advection velocity (equation \ref{eq:mag_trans_new_velocity}). 

As the magnetic field is generated azimuthally in this 2-D case, there is no component of heat-flow along magnetic field lines \cite{walsh2017}. Figure \ref{fig:diffTe_perp_wedge} shows the change in temperature at 1.7ns for the $\lambda=20\mu$m case due to both the magnetized perpendicular conduction ($\kappa_{\bot}$) and Righi-Leduc ($\kappa_{\wedge}$). The left side of the figure is the effect of turning on the magnetized perpendicular thermal conduction, while the right is the additional effect once Righi-Leduc is turned on (for physical consistency, the cross-gradient-Nernst advection of magnetic field is only included when Righi-Leduc is turned on \cite{2021}). The lowered value of $\kappa_{\bot}$ near the spike due to the magnetic field reduces the heat-flow into the spike, cooling the tip. Righi-Leduc re-directs heat-flow down from the tip of the spike, also cooling the tip. In this case the 2 terms are contributing a similar amount to the overall change in electron temperature, although this is not true for all cases analyzed.

The stabilization of RT spikes scales with the ablation velocity of the spikes \cite{betti2001}. Here we assume that the only process causing ablation is the electron heat-flow. The ablation velocity can be obtained by assuming that all of the electron energy going into the cold fuel is then ablated into the hot-spot \cite{betti2001}, giving $V_{abl} \sim \frac{\kappa_{\parallel}^c T_e^{5/2}}{L_T \rho}$ for the unmagnetized case. In this letter it is proposed to replace the unmagnetized conductivity $\kappa_{\parallel}^c$ with the magnetized $\kappa_{\bot}^c$ and introduce an additional Righi-Leduc heat-flow that reduces ablation at the spike tip. Therefore, the RT stabilization including magnetic fields is:

\begin{equation}
	V_{MHD} \sim \frac{(\kappa_{\bot}^c - \kappa_{\wedge}^c)T_e^{5/2}}{L_T \rho}
\end{equation}
i.e. the change in spike velocity due to magnetic fields is as given in equation \ref{eq:delta_V}. The composite function of different $\kappa^c$ coefficients is given in figure \ref{fig:kappa} for varying Hall Parameter, showing that the Righi-Leduc term dominates at low $\omega_e \tau_e$, but the suppressed thermal conduction has a greater impact for $\omega_e \tau_e>0.5$. 

A comparison of equation \ref{eq:delta_V} to the simulations is given in figure \ref{fig:scaling_c} with significant deviations from the scaling only when magnetic flux injection occurs. 

The impact of MHD on the spike heights for various wavelengths and initial perturbation sizes are shown in figure \ref{fig:del_h}. The impact of MHD continually increases until the spike is effectively stabilized, which allows for the magnetic field to detach from the spike. As the perturbation wavelength is decreased, the magnetic fields enhance the instability earlier in time, due to the $1/\lambda^2$ dependence of the magnetization from equation \ref{eq:wt}. However, the shorter wavelengths are also stabilized at an earlier time, which decreases the maximum impact of MHD. The large initial perturbation cases in figure \ref{fig:del_h} show that a $\lambda=10\mu$m spike has its propagation enhanced by $1.6\mu$m by 1.55ns, before the spike is stabilized. For the $\lambda=20\mu$m case, however, the spike (and magnetic flux) continue to grow all the way past 1.7ns, by which time the MHD has enhanced the spike growth by $3.5\mu$m. This demonstrates the importance of the electron magnetization scaling with $h\int h \delta t$ from equation \ref{eq:wt}.

In summary, a scaling for the enhancement of the non-linear ablative Rayleigh-Taylor instability due to self-generated magnetic fields has been presented, comparing favorably with 2-D extended-MHD results. 

The theory can be used to post-process capsule simulations to estimate the impact of magnetic fields. To demonstrate this, 2-D capsule simulations of HDC implosion N170601 are used with multi-mode shell thickness asymmetries. Previous work quantifying magnetic flux generation predicted peak flux for modes 40-60 \cite{walsh2021a}. Using the theory developed in this paper, the dependence of electron magnetization on mode number can now be estimated, with peak magnetization for modes 20-35. The peak Hall Parameter is calculated as up to 0.8, which results in $\Delta V_{MHD}/V_{abl} \approx 1$, i.e. significant reduction in stabilization due to the self-generated magnetic fields for those modes. Future work will also quantify how the self-generated fields modify hot-spot temperature and fusion yield; this is of particular interest for the recent ignition experiment on the National Ignition Facility \cite{PhysRevLett.129.075001,PhysRevE.106.025201,PhysRevE.106.025202}, which is expected to have generated more magnetic flux than previous experiments due to the larger hot-spot temperatures \cite{walsh2021a}.

The theory deviates from simulations once the magnetic field detaches from the spike, which occurs when the ablative stabilization becomes so strong that the spike begins to decrease in size. At this stage the magnetic flux loops are injected into the hot plasma, which reduces the impact on Rayleigh-Taylor growth but continues to magnetize the local electron population and may still modify ICF hot-spot performance.
	
	\section*{Acknowledgements}
	This work was performed under the auspices of the U.S. Department of Energy by Lawrence Livermore National Laboratory under Contract DE-AC52-07NA27344. 
	
	This document was prepared as an account of work sponsored by an agency of the United States government. Neither the United States government nor Lawrence Livermore National Security, LLC, nor any of their employees makes any warranty, expressed or implied, or assumes any legal liability or responsibility for the accuracy, completeness, or usefulness of any information, apparatus, product, or process disclosed, or represents that its use would not infringe privately owned rights. Reference herein to any specific commercial product, process, or service by trade name, trademark, manufacturer, or otherwise does not necessarily constitute or imply its endorsement, recommendation, or favoring by the United States government or Lawrence Livermore National Security, LLC. The views and opinions of authors expressed herein do not necessarily state or reflect those of the United States government or Lawrence Livermore National Security, LLC, and shall not be used for advertising or product endorsement purposes.
	
	\section*{Bibliography}
	
	\bibliographystyle{unsrt}

		\ifdefined\DeclarePrefChars\DeclarePrefChars{'’-}\else\fi

\end{document}